\documentclass[letterpaper,10pt,conference]{ifacconf}

\usepackage{amsmath} 
\usepackage{amssymb}  
\usepackage{mathtools}
\usepackage{graphicx}
\usepackage{natbib}        
\usepackage{caption,subcaption}
\usepackage{color}
\usepackage{enumerate}
\usepackage{epstopdf}
\usepackage{dsfont}
\usepackage[font=small,skip=4pt]{caption}

\usepackage{pgfplots}
\pgfplotsset{width=10cm, compat=1.9, legend style={font=\footnotesize}}
\tikzset{%
dot/.style={circle, fill=black, minimum size=4pt, inner sep=0pt, outer sep=-1pt},
hdot/.style={circle, fill=white, minimum size=4pt, inner sep=0pt, outer sep=-1pt},
}
\newlength\fheight 
\newlength\fwidth 
\usepackage{circuitikz}
\usepgflibrary{arrows}

\definecolor{new}{RGB}{0,0,0}
\definecolor{ajg}{RGB}{0,0,0}
\definecolor{gft}{RGB}{0,0,0}
\definecolor{mst}{RGB}{0,0,0}
\definecolor{fb} {RGB}{0,0,0}
\definecolor{tp} {RGB}{0,0,0}

\usepackage{tikz}
\usetikzlibrary{calc, backgrounds}
\usetikzlibrary{shapes.geometric, arrows}

\tikzstyle{process}  = [rectangle, rounded corners, minimum width=2cm, minimum height=1cm,text centered, text width=2cm, draw=black, fill=red!20]

\tikzstyle{decision} = [rectangle, minimum width=2cm, minimum height=1cm, text centered, text width=2.5cm, draw=black, fill=orange!20]

\tikzstyle{scenario} = [rectangle, rounded corners, minimum width=2cm, minimum height=0.5cm, text centered, text width=3cm, draw=black, fill=pink!20]

\tikzstyle{arrow}= [thick, ->, >=stealth]

\tikzstyle{system} = [rectangle, rounded corners, minimum width=4cm, minimum height=4cm, text centered, text width=4cm, draw=black, fill=red!40,fill opacity = 0.2]

\tikzstyle{controller} = [rectangle,rounded corners,minimum width=1cm,minimum height = 1cm,text centered,text width=1cm,draw=black,fill=green!50,fill opacity = 0.2]

\tikzstyle{detector} = [rectangle,rounded corners,minimum width=1cm,minimum height = 1cm,text centered,text width=1cm,draw=black,fill=yellow!50,fill opacity = 0.2]

\newtheorem{proposition}{Proposition}

\newtheorem{assumption}{Assumption}
\newtheorem{remark}{Remark}

\newcommand{\NN}{{\mathcal{N}}}

\newcommand{\commentout}[1]{}

\begin{document}

\begin{frontmatter}
		
		\title{Distributed watermarking for secure control of microgrids under replay attacks
		}
		
		\thanks[footnoteinfo]{This work has been partially supported by European Union's Horizon 2020 research and innovation programme under grant agreement No 739551 (KIOS CoE). This work has also been conducted as part of: i) the research project  {\em Stability and Control of Power Networks with Energy Storage} (STABLE-NET)  which is funded by the RCUK Energy Programme (contract no: EP/L014343/1); ii)~the Swiss National Science Foundation	under the COFLEX project (grant number 200021\_169906).}
		
		\author[First]			{Alexander J. Gallo} 
		\author[Second]			{Mustafa S. Turan} 
		\author[Third]			{Francesca Boem}
		\author[Second]			{Giancarlo Ferrari-Trecate}
		\author[First,Fourth]	{Thomas Parisini}
		
		\address[First]{
			Imperial College London, London, UK. (emails: alexander.gallo12@imperial.ac.uk,t.parisini@gmail.com).}
		\address[Second]{
			\'Ecole Polytechnique F\'ed\'erale de Lausanne (EPFL), Switzerland. (e-mails: \{mustafa.turan, giancarlo.ferraritrecate\}@epfl.ch)}
		\address[Third]{
			University College London, UK. (e-mail: f.boem@ucl.ac.uk)}
		\address[Fourth]{
			University of Trieste, Italy and KIOS Research and Innovation Centre of Excellence, University of Cyprus.}

	\begin{abstract}
        The problem of replay attacks in the communication \textcolor{ajg}{network} between Distributed Generation 			Units \textcolor{ajg}{(DGUs)} of a DC microgrid is examined. \textcolor{fb}{The DGUs are regulated through 			a hierarchical control architecture, and \textcolor{ajg}{are networked}
        to achieve secondary control objectives.} Following analysis of the detectability of replay attacks 
		\textcolor{fb}{by} a distributed monitoring scheme previously proposed, the need for a watermarking signal 			is identified. Hence, 
		conditions are given on the watermark in order to guarantee detection \textcolor{fb}{of replay attacks}, 		 and \textcolor{ajg}{such}
        a 
        signal is designed. 
        Simulations are then presented to demonstrate the effectiveness of the technique.
	\end{abstract}

\end{frontmatter}

\section{Introduction}\label{ch:Intro}
\subsection{Motivation and state of the art}
The interest in islanded microgrids has spiked over the past few years, following the increased penetration of renewable energy sources within the electrical network
\textcolor{ajg}{as they can offer benefits compared to traditional grids,}
\textcolor{new}{such as improving reliability, help avoiding overload problems, and allowing local power quality regulation, among others}
\citep{meng2017review}.
Specifically, as many of the adopted generation, storage, and end-user equipment are based on DC technology, 
DC microgrids (DCmGs) have attracted a lot of research activity, as can be seen from the recent survey \citep{meng2017review}. In order to guarantee the stable and efficient behavior of the microgrid, hierarchical control architectures have been proposed \citep{Guerrero}, in which a decentralized primary layer ensures voltage, current and power stability \citep{tucci2016decentralized,zhao2015distributed}, and secondary and tertiary control layers\textcolor{new}{, then, }
offer additional properties, such as power quality regulation, load 
sharing, and ensuring overall coordination and optimization \citep{DePersis,tucci2017consensus,zhao2015distributed}. It has been shown 
that\textcolor{ajg}{,} in order to achieve some of the objectives, the secondary and tertiary controllers require the support of a communication network \textcolor{new}{between Distributed Generation Units (DGUs)}
\textcolor{mst}{\citep{Ferrari, Carli}}.

The integration of communication networks within control systems has exposed them to the possibility of being tampered by malicious agents\textcolor{gft}{, injecting false} 
information within the control loop, thus altering \textcolor{ajg}{their behavior} 
\citep{cardenas2008secure,cheng2017guest}. 
Given the possibility of attacks, it has been recognized as necessary to introduce monitoring structures capable of evaluating whether operations are running as normal or not \textcolor{new}{(see papers in \citep{sandberg2015cyberphysical,cheng2017guest}).}

\textcolor{ajg}{Among several types of attacks, during \textit{replay attacks} an attacker is able to record data transmitted over a communication network, and then \textit{replay} it, replacing actual communication signals with buffered data. This class of attacks has been shown to be \textcolor{tp}{particularly difficult to detect for common monitoring schemes,}
as they present the same statistics as the nominal behavior \citep{mo2015physical}. Several techniques have been proposed in order to detect this class of attack, often by altering the characteristics of the system through the addition of a \textit{watermark} \citep{mo2015physical,ferrari2017detection}.}
\textcolor{new}{These methods borrow the idea of \textit{watermarking} from the multimedia industry \citep{hartung1999multimedia}. where data is embedded in order to prevent unauthorized reproduction of media content.}
In \citep{mo2015physical}, a \textcolor{ajg}{time-varying watermark} is added to the input signal \textcolor{mst}{of a system,} in order to alter the characteristic statistics of the steady state, thus allowing the monitoring scheme to detect the presence of an attack. In \citep{ferrari2017detection}, the watermark signal is added directly to the sensor measurements communicated to the monitoring scheme and controller. 
\textcolor{black}{Other techniques have been proposed to counteract stealthy data injection attacks. For example \textcolor{fb}{in} \citep{miao2017coding}, the proposed strategy encodes the sensor measurements through a \textit{"coding matrix"}, assumed to be unknown by the attacker, in order to prevent stealthy data injection.}

All of the methods proposed to detect replay attacks rely on the possibility of implementing the monitoring scheme centrally. This, however, is not desirable for DCmGs, \textcolor{mst}{as all distributed generation units (DGUs) would need to communicate with a central point, which would increase the communication cost.} 
In this preliminary work, we propose a distributed watermarking technique for the monitoring of microgrids.

\textcolor{fb}{Recently, a few works have examined attacks on microgrids, although not considering replay attacks.}
\textcolor{ajg}{The susceptibility of the secondary control objectives to jamming attacks on the communication between DGUs is shown in \citep{danzi2016impact}, while in \citep{gallo2018distributed} we have proposed a distributed monitoring scheme to detect the presence of attackers in the communication network,} developed off the preliminary work in \citep{boem2017distributed}.

\subsection{Objectives and contributions}\label{subch:Objectives}

In this paper, we consider a DCmG regulated as in \citep{tucci2017consensus}, and monitored as in \citep{gallo2018distributed}. We introduce a distributed watermarking scheme which allows for the detection of replay attacks in the communication network. The main objectives that \textcolor{new}{the proposed watermarking technique} fulfills are:
\begin{enumerate}[a.]
	\item Enhance the monitoring scheme in \citep{gallo2018distributed}, in order to detect replay attacks;
	\item Be distributed, \textcolor{gft}{running attack detectors at each DGU location and using 
    the same communication network required for secondary control in \citep{tucci2017consensus}};
	\item Ensure the main goals of the primary and secondary controllers, i.e. voltage regulation and current 	  sharing, respectively, are \textcolor{gft}{not compromised by watermarking};
	\item Require a watermark which is not easily identifiable by a resourceful attacker.
\end{enumerate}

\textcolor{gft}{Objective (a) is motivated by the fact that the monitoring scheme in \citep{gallo2018distributed} is vulnerable to replay attacks, as shown in Section~\ref{ch:CyberAtk_Mtr}.} We also \textcolor{fb}{derive} detectability conditions, fundamental for the design of the watermark, which must be fulfilled in order to guarantee detection. \textcolor{gft}{This analysis will allow us to design} a preliminary watermark which enables the detection of replay attacks.

\subsection{Paper structure}
The rest of the paper is structured as follows. In Section \ref{ch:DCmG} we introduce the model of the DCmG which is considered. Section \ref{ch:CyberAtk_Mtr} is dedicated to the introduction of the monitoring scheme in \citep{gallo2018distributed}, as well as the definition of the replay attack, and the demonstration that it is stealthy to the scheme \textcolor{new}{in \citep{gallo2018distributed}}. In Section \ref{ch:Detect_RepAtk} we analyze the detectability properties of a generic watermarking scheme, and propose a preliminary watermark design. Some simulation results are shown in Section \ref{ch:Sim}, \textcolor{fb}{where} the proposed watermark is shown to be effective at detecting replay attacks. {Finally, in Section \ref{ch:Concl}, we summarize the results, and define some future research directions we believe to be of interest.

\subsection{Notation} In the paper, the operator $|\cdot|$ applied to a set determines its cardinality, while used with matrices or vectors it defines their component-by-component absolute value. The operator $\|\cdot\|$ is used to define the matrix norm. In general, in this paper inequalities are considered component-by-component. The operators $\lceil \,\cdot\, \rceil$ and $\lfloor \,\cdot\, \rfloor$ define, respectively, the ceiling and floor functions.

\section{Characterization of DC Microgird}\label{ch:DCmG}

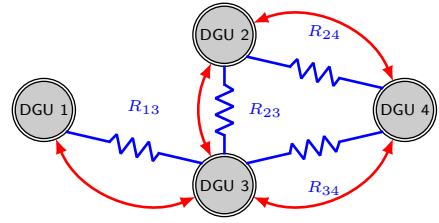
\begin{figure}
	\centering
	{\tiny
		\ctikzset{bipoles/length=0.8cm}
		\tikzstyle{every node}=[font=\sffamily\tiny,  minimum size=.8cm, inner sep=.8, line width=0.5pt]	\begin{circuitikz}[american currents, scale=0.8, line width=1pt]
			\draw (1,1) node(a) [circle, draw=black, double,
			fill=black!20] {DGU 1};
			\draw (7,1) node(b)  [circle, draw=black, double,fill=black!20] {DGU 4};
			\draw (4,-.25) node(c)  [circle, draw=black, double, fill=black!20] {DGU 3};
			\draw (4,2.25) node(d)  [circle, draw=black, double,  fill=black!20] {DGU 2};

			\draw[blue] (b.south west)  to [R, l={$R_{34}$}, color=blue] (c.north east);  
			
			\draw[blue] (d.south east) to [R, l=$R_{24}$, color=blue] (b.north west);

			\draw[blue] (d.south) to [R, l=$R_{23}$, color=blue] (c.north);
			
			\draw[blue] (a.south east) to [R, label=$R_{13}$, color=blue] (c.north west);

			\draw[latex-latex, red] (c) to [bend right=45] (b);
			\draw[latex-latex, red] (c) to [bend left=45]  (a);
			\draw[latex-latex, red] (d) to [bend left=45]  (b);
			\draw[latex-latex, red] (d) to [bend right=25] (c);		
	\end{circuitikz}}
	\caption{Graph of DCmG. The blue lines represent the power lines connecting the DGUs, and the red arrows represent the communication graph. 
	}
	\label{fig:mG_Struct}
\end{figure}

{In the following, we consider a DCmG 
\textcolor{new}{composed of} $N$ distributed generation units interconnected by power lines. Each DGU is modeled as in \citep{tucci2016decentralized}, where \textcolor{gft}{the} interconnection power lines are taken to be purely resistive\textcolor{gft}{.} The DCmG modeled in this framework can therefore be represented as an undirected graph, where nodes represent the DGUs, edges denote power lines, and \textcolor{ajg}{edge weights} are given by the conductance of the lines \textcolor{gft}{(see Fig~\ref{fig:mG_Struct}).} DGUs are defined as the \textcolor{gft}{cascade of a} DC voltage source, a DC/DC converter, and a RLC filter connecting it to other DGUs at a point of common coupling (PCC). DGUs have local loads, represented as known currents $I_{Li}$. The dynamics of each DGU are characterized in state space as follows:
	\begin{equation}\label{eq:DGUi}
	\begin{split}
	&\dot{x}_{[i]}(t) = A_{ii}x_{[i]}(t) + B_iu_{[i]}(t) + G_i\alpha_{[i]}(t) \\
	& \qquad\qquad\qquad + M_id_{[i]}(t) + \xi_{[i]}(t) + w_{[i]}(t)\\
	&y_{[i]}(t) = C_ix_{[i]}(t)	+ \rho_{[i]}(t)
	\end{split}~~,
	\end{equation}
where $x_{[i]} = [V_i,I_{ti},v_{[i]}]^{\top}$ is the state of the DGU, $V_i$ is the voltage at the \textcolor{ajg}{PCC,} $I_{ti}$ is the terminal current from the converter, and $v_{[i]}$ is \textcolor{gft}{a scalar state needed \textcolor{ajg}{to include} an integrator action in the primary loop}. $u_{[i]}$ and $\alpha_{[i]}$ are respectively the primary and secondary input,\textcolor{ajg}{where} the latter is defined in \eqref{eq:cons}. Vector ${d_{[i]} = [I_{Li},V_{ref,i}]^{\top}}$ is an external known input to DGU~$i$, where $V_{ref,i}$ is a voltage reference for $V_{i}$. ${\xi_{[i]} = \sum_{j\in\mathcal{N}_i}A_{ij}x_{[j]}}$ is a vector modeling the interconnection with other DGUs, where $\mathcal{N}_i\subset\mathcal{N}\equiv\{ 1,\cdots,N\}$ is the set of \textcolor{mst}{the neighbors of DGU~$i$}, i.e. the DGUs that are interconnected to DGU~$i$ through power lines.}
{Vectors $w_{[i]}$ and $\rho_{[i]}$ model unknown state and measurement {noises}, respectively. \textcolor{new}{We assume the following:}
	\begin{assumption}\label{ass:Bounds}
		The unknown process noise $w_{[i]}$ and measurement noise $\rho_{[i]}$ are bounded for all time:
		\begin{equation}
		|w_{[i]}(t)|\leq \bar{w}_{[i]}, |\rho_{[i]}(t)|\leq \bar{\rho}_{[i]}, \forall t,
		\end{equation}        
        where $\bar{w}_{[i]}$ and $\bar{\rho}_{[i]}$ are known. $\hfill \triangle$
\end{assumption}}

The primary input $u_{[i]} = V_{ti} = K_iy_{[i]}$ is given by a decentralized feedback controller \textcolor{ajg}{\citep{tucci2016decentralized},} \textcolor{new}{where $K_i$ is a matrix designed according to \citep{tucci2016decentralized},} and guarantees voltage stability and reference tracking \textcolor{gft}{in the whole \textcolor{ajg}{DCmG}}. The secondary control input $\alpha_{[i]}$ allows for current to be shared among DGUs, and is the result of the following consensus-based protocol:
	\begin{equation}\label{eq:cons}
	\dot{\alpha}_{[i]}(t) = -\sum_{j \in \NN_i} \left[0\,k_I\,0\right] \left(\frac{y_{[i]}(t)}{I_{ti}^s} - \frac{y^c_{[j,i]}(t)}{I_{tj}^s} \right),
	\end{equation}
where the scalars $I_{ti}^s>0,\forall i \in \mathcal{N}$ are \textcolor{gft}{design parameters} which allow \textcolor{new}{the current to be shared among DGUs} at different rates. Scalar $k_I>0$ is the consensus weight, common to all DGUs. Vector $y_{[j,i]}^c(t)$ represents the output measurement which DGU~$j$ transmits to DGU~$i$, as defined in \eqref{eq:yComm} \textcolor{gft}{below}. The matrices in \eqref{eq:DGUi} are defined as in \citep{tucci2016decentralized} \textcolor{new}{and are summarized in the Appendix.}

{In order to operate the consensus-based secondary controller described in \eqref{eq:cons}, it is necessary to introduce a communication network connecting different DGUs\textcolor{new}{, assumed to have the same topology as the DCmG}. For ease of analysis, in this preliminary paper, we introduce the following assumption:
\begin{assumption}
The communication network is ideal, i.e. information is exact \textcolor{ajg}{and} without time delays.
\textcolor{ajg}{Hence} communicated data \textcolor{ajg}{is:
\begin{equation}\label{eq:yComm}
y_{[i,j]}^c(t) = y_{[i]}(t)~~,
\end{equation}
i.e.}
is equal to the measurement vectors. $\hfill \triangle$
\end{assumption}
The introduction of a communication network 
\textcolor{fb}{exposes the system to malicious attacks.} 
In the following section we describe how we model the attack strategy, and define the type of attack motivating this work.
}

\section{Cyber-attacks and monitoring scheme}\label{ch:CyberAtk_Mtr}

\textcolor{gft}{Our goal is} to equip all DGUs with a monitoring scheme\textcolor{gft}{, allowing} to check whether the information received from each neighbor is corrupted by an attack or not. In order to formally introduce the attack in the communication value $y_{[i,j]}^c(t)$, we redefine \eqref{eq:yComm} as the following:
\begin{equation}\label{eq:yCommAtk}
y_{[i,j]}^c(t) = y_{[i]}(t) + \beta_{ij}(t-T_a)\phi_{i,j}(t)
\end{equation}
where $\phi_{i,j}(\cdot)$ is the attack function - designed by the attacker according to its objectives and available resources - and 
\textcolor{new}{$\beta_{ij}(t-T_a)$ is an activation function, which is $0$ for $t<T_a$, and $1$ otherwise, {where $T_a$ is the time} at which the attack begins.}

In order to detect the action of an attacker, a distributed attack detection scheme based on multiple Unknown Input Observers (UIOs) has been introduced \citep{gallo2018distributed}. This monitoring strategy is capable of detecting whether the communication is subject to an attack, based on limited knowledge of the neighbors' dynamics, and on information regarding the bounds on the disturbance in Assumption \ref{ass:Bounds}. We will now briefly summarize the monitoring strategy \textcolor{new}{in \citep{gallo2018distributed}.}

\subsection{Monitoring strategy}
In \textcolor{ajg}{the scheme proposed in} \citep{gallo2018distributed}, 
\textcolor{new}{we have introduced a monitoring scheme based on UIOs, where}
each DGU estimates the state of each of its neighbors. The error between the estimate and the received measurement\textcolor{fb}{s vector} is then compared to a time-varying threshold, designed based on the bounds 
\textcolor{new}{on the noises defined} in Assumption \ref{ass:Bounds}, \textcolor{ajg}{to determine} whether the communication network is \textcolor{new}{secure} or not. In the remainder, we will present the estimation scheme in DGU~$i$ for the state of its neighbor $j$. The UIO framework is used to reduce the information transmitted between DGUs, as any \textcolor{ajg}{communicated data} may be \textcolor{ajg}{exposed to attacks.}

In order to exploit the UIO estimator, we rewrite the dynamics of DGU~$j$ as follows:
\begin{equation}\label{eq:DGUjDist}
\begin{split}
&\dot{x}_{[j]}(t) = A_{Kj}x_{[j]}(t) + \bar{E}_j\bar{d}_{[j]}(t) + \tilde{w}_{[j]}(t) \\
&y_{[j]}(t) = C_jx_{[j]}(t) + \rho_{[j]}(t)
\end{split}
\end{equation}
where $A_{Kj} = A_{jj} + B_jK_j$, $\tilde{w}_{[j]}(t) = w_{[j]}(t)  + B_jK_j\rho_{[j]}(t)$, \textcolor{mst}{and}
$\bar{E}_j\bar{d}_{[j]}$ represents the inputs to DGU~$j$ which are unknown to the UIO in DGU~$i$. 
Specifically, $\bar{d}_{[j]} \textcolor{gft}{= \widehat{E}_j\hat{d}_{[j]}}$ is a linear combination of the vector of variables:
\[
\hat{d}_{[j]} = \left[
d_{[j]}^\top(t) , \alpha_{[j]}(t) , x_{[k_1]}^\top(t), \dots , x_{[k_{|\NN_j|}]}^\top(t)
\right]^\top,
\]
while $\bar{E}$ is \textcolor{ajg}{full column rank and} derived from matrices in \eqref{eq:DGUi}, \textcolor{gft}{and} is defined in \citep{gallo2018distributed} \textcolor{gft}{along with $\widehat{E}_j$}.

\textcolor{new}{The state of the UIO and estimate} are \textcolor{gft}{\citep{chen1996design}}:
\begin{equation}\label{eq:EstUIOj}
\begin{split}
&\dot{z}_{[j,i]}(t) = F_jz_{[j,i]}(t) + S_jB\bar{u}_{[j]}(t) + \widehat{K}_jy^c_{[j,i]}(t) \\
&{\hat{x}}_{[j,i]}(t) = z_{[j,i]}(t) + H_jy_{[j,i]}^c(t)
\end{split}
\end{equation}
where the matrices are defined as in \citep{gallo2018distributed}, and are such that $F_j$ is Hurwitz stable, and $S_j\bar{E}_j = 0$. The residual is defined as $r_{[j,i]}(t) = y^c_{[j,i]}(t) - \hat{x}_{[j,i]}(t)$, and, given stability of \eqref{eq:EstUIOj} and Assumption \ref{ass:Bounds}, is bounded, \textcolor{ajg}{i.e.}
\begin{equation}\label{eq:atkDetTst}
\left|r_{[j,i]}(t)\right| \leq \bar{r}_{[j,i]}(t)
\end{equation}
holds for all $t\geq 0$, where 
\begin{equation}\label{eq:resBound}
\begin{split}
\bar{r}_{[j,i]}(t) = \bar{e}_{[j,i]}(t) + \bar{\rho}_{[j]}.
\end{split}
\end{equation}

\textcolor{ajg}{is the detection threshold.} \textcolor{gft}{In \eqref{eq:resBound} ,} $\bar{e}_{[j,i]}(t)$ is the time-varying bound on the estimation error $e_{[j,i]}(t) = x_{[j]}(t) - \hat{x}_{[j,i]}(t)$, defined as:
\begin{equation}\label{eq:estErrBound}
\begin{split}
\bar{e}&_{[j,i]}(t) = \kappa e^{-\mu t} \left[\bar{e}_{[j,i]}(0) + |H_j|\bar{\rho}_{[j]} \right] + 		|H_j|\bar{\rho}_{[j]} \\
&+ \int_0^t \kappa e^{-\mu (t-\tau)} \left[|S_j|\bar{w}_{[j]} + |S_jB_jK_j - \widehat{K}_j|\bar{\rho}_{[j]}  \right]d\tau
\end{split}
\end{equation}
where scalars $\kappa,\mu > 0$ are such that $\|e^{F_jt}\|\leq\kappa e^{-\mu t}$. This bound, for suitably defined $\bar{e}_{[j,i]}(0)$, guarantees that $|e_{[j,i]}(t)|\leq\bar{e}_{[j,i]}(t), \forall t \geq0$. Given these bounds on the estimation error and residual, an attack is detected if \textcolor{ajg}{\eqref{eq:atkDetTst}, used as a detection test, is not satisfied.}

The detectability conditions of this scheme have been studied in \citep{gallo2018distributed}. 
\textcolor{tp}{In the following section we show that it may fail to detect \textit{replay attacks}.}


\subsection{Replay attack}

\textcolor{tp}{As anticipated in the introduction, we focus specifically on the class of replay attacks, as even without any knowledge of the system model and much computational resources, they can be undetectable \citep{mo2015physical}.}
In the scenario we are considering, a malicious agent is able to eavesdrop
the communication \textcolor{new}{link} between DGU~$j$ and DGU~$i$, \textcolor{ajg}{from} some unknown time $t = T_0$, and thus to \textcolor{gft}{start storing} this data in a memory buffer \textcolor{gft}{up to a time $T_a >T_0$}. At time $t = T_a$, \textcolor{gft}{the attacker} starts injecting the following attack function into the communicated signal \eqref{eq:yCommAtk}:
\begin{equation}
\phi_{[j,i]}(t) = - y_{[j,i]}^c(t) + y^c_{[j,i]}(t-nT),
\end{equation}
i.e. it replaces the current 
transmitted measurements of the output of DGU~$j$ \textcolor{gft}{with past recorded measurements.} The integer $n = \left\lceil {(t-T_a)}/{T} \right\rceil$ \textcolor{fb}{represents} the periodicity of the signal the attacker injects, \textcolor{ajg}{where $T\leq T_a-T_0$ is the period of the repeated data, as decided by the attacker.}

\textcolor{tp}{These attacks are particularly deceptive if data is recorded when the state of the system is {in a} quasi stationary {r\'{e}gime}. In 
Prop.~\ref{prop:stealthRep}, we give a preliminary result showing a sufficient condition for the attack to be stealthy.}

\begin{proposition}[Stealthy Replay Attacks]\label{prop:stealthRep}
	\textcolor{ajg}{If 
    \begin{equation}\label{eq:hypStealth}
    |e_{[j,i]}^a(T_a)| \leq \bar{e}_{[j,i]}(T_a)
    \end{equation}
    \textcolor{tp}{is satisfied,} where $e_{[j,i]}^a(T_a) = x(T_a-T) - \hat{x}_{[j,i]}(T_a)$, then \textcolor{tp}{inequality} \eqref{eq:atkDetTst} holds for all $t\in[T_a,T_a+T)$\textcolor{tp}{, and the attack will not be detected.} \textcolor{tp}{$\hfill \square$}}
\end{proposition}
\begin{pf}
\textcolor{new}{Given time $T_a$ and period of stored data $T$, for $t\geq T_a$, the residual $r_{[j,i]}(t) = \hat{y}^c_{[j,i]}(t) - \hat{x}_{[j,i]}(t)$ is
$$r_{[j,i]}(t) = e^a_{[j,i]}(t) + \rho_{[j]}(t-nT) .$$
Note that $e^a_{[j,i]}(t)$ is the estimation error with the replay attack. To find the value of $e^a_{[j,i]}(t)$, we analyze its dynamics derived from \eqref{eq:DGUi} and \eqref{eq:EstUIOj}. Hence, for $t\in[T_a,T_a+T)$ (i.e. $n = 1$), the dynamics of the estimation error under attack are:
\begin{equation}\label{eq:estErrAtk1}
\begin{split}
	\dot{e}_{[j,i]}^a(t) = F_j e_{[j,i]}^a(t) + S_j\tilde{w}_{[j]}(t-T) - 		\widetilde{K}_j\rho_{[j]}(t-T) \\
	- H_j\dot{\rho}_{[j]}(t-T).
\end{split}
\end{equation}
The solution to \eqref{eq:estErrAtk1} is given by:
\begin{equation}
\begin{split}
	e_{[j,i]}^a(t) = e&^{F_j(t-T_a)}e_{[j,i]}^a(T_a) + \int_{T_a}^t e^{F_j(t-\tau)} \left[S_j\tilde{w}_{[j]}(\tau-T)+\right. \\
	& \left.- \widetilde{K}_j\rho_{[j]}(\tau-T) - H_j \dot{\rho}_{[j]}(\tau-T)    \right] d\tau\\
\end{split}
\end{equation}
which, by  use of integration by parts, is:
\begin{equation}\label{eq:estErrAtkSln}
\begin{split}
	&e_{[j,i]}^a(t) = e^{F_j(t-T_a)}\left(e_{[j,i]}^a(T_a) +H_j \rho_{[j]}(T_a-T)\right) +\\ 
	&- H_j\rho_{[j]}(t-T) + \int_{T_a}^t e^{F_j(t-\tau)} \left[S_j{w}_{[j]}(\tau-T) + \right. \\
	&\left.+ (S_jB_jK_j-\widehat{K}_j)\rho_{[j]}(\tau-T) \right] d\tau ,
\end{split}
\end{equation}
where $e_{[j,i]}^a(T_a)= x_{[j]}(T_a-T)-\hat{x}_{[j,i]}(T_a)$ is the estimation error at the start of the attack. Let us also note that the upper bound on the estimation error $e_{[j,i]}(t)$ is given as:
	\begin{equation}\label{eq:estErrBoundTa}
	\begin{split}
	&\bar{e}_{[j,i]}(t) =  \kappa e^{-\mu(t-T_a)} \left[\bar{e}_{[j,i]}(T_a) + \left|H_j\right|\bar{\rho}_{[j]} \right] + 		\left|H_j\right|\bar{\rho}_{[j]} \\
	&+ \int_{T_a}^t \kappa e^{-\mu (t-\tau)} \left[\left|S_j\right|\bar{w}_{[j]} + \left|S_jB_jK_j - \hat{K}_j\right|\bar{\rho}_{[j]}  \right]d\tau.
	\end{split}
	\end{equation}
	We observe that all the terms in (\ref{eq:estErrAtkSln}), apart from $e^a_{[j,i]}(T_a)$, are bounded by the corresponding terms in (\ref{eq:estErrBoundTa}). Hence, 
	it is sufficient that $|e^a_{[j,i]}(T_a)|\leq\bar{e}_{[j,i]}(T_a)$, where $\bar{e}_{[j,i]}(T_a)$ is defined in \eqref{eq:estErrBound}, for the detection threshold \eqref{eq:atkDetTst} to not be violated for all $t \in [T_a,T_a+T)$, thus proving the Proposition. $\hfill \blacksquare $}	
\end{pf}

\begin{remark}
	The objective of the malicious agent executing a replay attack is to hide any change in operating conditions in DGU~$j$ from DGU~$i$, i.e. the changes caused by a load change in DGU~$j$ or one of its neighbors. By doing so, \textcolor{fb}{the attack}
is able to alter the equilibrium which is reached through consensus, thus impeding current sharing, or it may even be able to make it impossible to reach consensus. \textcolor{new}{On the other hand, it is clear to see that, as long as steady state is maintained, it will not have any effect on the network.}
\end{remark}

In the following we will present a detection strategy based on \textit{watermarking} to detect the presence of replay attacks in the communication network.

\section{Cyber-attack detection method}\label{ch:Detect_RepAtk}

\textcolor{ajg}{To} make a replay attack detectable, \textcolor{ajg}{similar to} the intuition behind \textcolor{ajg}{sensor} watermarking \citep{ferrari2017detection}, we add a time varying signal $\Delta_{[i,j]}(t)$ to the measurements communicated from DGU~$i$ to DGU~$j$. 
For this preliminary work the following \textcolor{ajg}{is assumed to not bias the performance of the consensus scheme (see objective (c) in Section~\ref{subch:Objectives})}:
\begin{assumption}\label{ass:knwWtmrk}
	Watermark $\Delta_{[i,j]}(t)$ added to $y_{[i,j]}(t)$ is known exactly by both DGU~$i$ and DGU~$j$ for all $t$. \textcolor{tp}{$\hfill \triangle$}
\end{assumption}

\subsection{Watermark signal}
With the addition of the watermark, the
communicated measurement \eqref{eq:yComm} becomes:
\begin{equation}\label{eq:yCommWtmk}
y_{[i,j]}^c(t) = y_{[i]}(t) + \Delta_{[i,j]}(t).
\end{equation}

Given Assumption \ref{ass:knwWtmrk}, once $y^c_{[j,i]}(t)$ is received by DGU~$i$, the known watermark is subtracted, as to achieve exact consensus \textcolor{new}{it is necessary that the watermark be absent.} \textcolor{ajg}{It is clear to see that the decoded information at DGU~$i$, $\hat{y}_{[j,i]}$, is the measurement of DGU~$j$: 
$$\hat{y}_{[j,i]}^c(t) = y_{[j,i]}^c(t) - \Delta_{[j,i]}(t) = y_{[j]}(t),$$ 
which is then used} in the dynamics of the secondary input \eqref{eq:cons}, as well as in the computation of the estimates \eqref{eq:EstUIOj} and \textcolor{ajg}{in the evaluation of the residual in inequality \eqref{eq:atkDetTst}.}

\textcolor{ajg}{Analyzing the effect of the watermark on the UIO estimators, it appears evident that}
under normal operating conditions, the value of 
$\hat{y}^c_{[j,i]}(t)$ will be the same as if the watermark weren't present. Hence, given analysis in \citep{gallo2018distributed}, the residual 
$r_{[j,i]}(t) = \hat{y}_{[j,i]}^c(t) - \hat{x}_{[j,i]}(t)$
does not exceed its bound, \textcolor{ajg}{avoiding false alarms, as \eqref{eq:atkDetTst} always holds.}
\textcolor{tp}{We now} analyze the residual \textcolor{ajg}{under replay attack.} 

\subsection{Detectability analysis}
For $t\geq T_a$, as previously mentioned, the information received by DGU~$i$ will be the buffered measurement $y_{[j,i]}^c(t-nT)$, and \textcolor{ajg}{hence decoded data is:} 
\begin{equation}
\hat{y}_{[j,i]}^c(t) = 
\textcolor{new}{y_{[j,i]}^c(t-nT) - \Delta_{[j,i]}(t) =}
y_{[j,i]}(t-nT) + \delta_{[j,i]}(t)
\end{equation}
where $\delta_{[j,i]}(t)$ is defined as
\begin{equation}\label{eq:deltaDef}
\delta_{[j,i]}(t) \coloneqq \Delta_{[j,i]}(t-nT) - \Delta_{[j,i]}(t).
\end{equation}

We also redefine the state estimation error in the presence of the watermark as $\epsilon^a_{[j,i]}(t) = x_{[j]}(t-T)-\hat{x}_{[j,i]}(t)$, and note that:
\begin{equation}\label{eq:errTaWM}
\epsilon^a_{[j,i]}(T_a) = e_{[j,i]}^a(T_a) - H_j\delta_{[j,i]}(T_a),
\end{equation}

To verify detectability conditions of the replay attack, it is necessary to analyze the residual under replay attacks, which must then be compared with the detection threshold $\bar{r}_{[j,i]}(t)$. Given \textcolor{new}{time instance} $T_a$ and \textcolor{new}{period of stored data} $T$, for $t\geq T_a$ residual \textcolor{new}{$r_{[j,i]}(t) = \hat{y}^c_{[j,i]}(t) - \hat{x}_{[j,i]}(t)$} is:
$$r_{[j,i]}(t) = \epsilon^a_{[j,i]}(t) + \rho_{[j]}(t-nT) + \delta_{[j,i]}(t).$$
For the value of $\epsilon^a_{[j,i]}(t)$, we analyze its dynamics starting from \eqref{eq:DGUi} and \eqref{eq:EstUIOj}.
\textcolor{gft}{In} the first period over which the replay attack is active, i.e. for $t\in[T_a,T_a+T)$, the dynamics of the estimation error are:
\begin{equation}\label{eq:estErrAtk}
\begin{split}
\dot{\epsilon}_{[j,i]}^a(t) = F_j \epsilon_{[j,i]}^a(t) + S_j\tilde{w}_{[j]}(t) - \widetilde{K}_j\left(\rho_{[j]}(t) +\delta_{[j,i]}(t)\right) + \\ 
- H_j\left(\dot{\rho}_{[j]}(t)+\dot{\delta}_{[j,i]}(t)\right).
\end{split}
\end{equation}
The explicit solution of these dynamics are:
\textcolor{new}{\begin{equation}
\begin{split}
&\epsilon_{[j,i]}^a(t) = e^{F_j(t-T_a)}\epsilon_{[j,i]}^a(T_a) +  \\
&\int_{T_a}^t e^{F_j(t-\tau)} \left[S_j\tilde{w}_{[j]}(\tau-T) - \widetilde{K}_j\left(\rho_{[j]}(\tau-T)+\delta_{[j,i]}(\tau)\right) \right.\\
& \qquad\qquad\qquad\qquad \left.- H_j \left( \dot{\rho}_{[j]}(\tau-T) + \dot{\delta}_{[j,i]}(\tau) \right)   \right] d\tau
\end{split}
\end{equation}
which, by use of integration by parts and substituting in equation \eqref{eq:errTaWM}, is:}
\begin{equation}\label{eq:errRepAtk}
\begin{split}
\epsilon_{[j,i]}^a&(t) = e^{F_j(t-T_a)}\left(\left[e_{[j,i]}^a(T_a) - H_j\delta_{[j,i]}(T_a)\right] +\right.\\
+&\left.H_j \rho_{[j]}(T_a-T) + H_j\delta_{[j,i]}(T_a)\right)  - H_j\left(\rho_{[j]}(t-T)+\right.\\
+&\left.\delta_{[j,i]}(t)\right) + \int_{T_a}^t e^{F_j(t-\tau)} \left[S_j{w}_{[j]}(\tau-T) + \right.\\
+&\left.(S_jB_jK_j-\widehat{K}_j)\rho_{[j]}(\tau-T) - \widehat{K}_j {\delta}_{[j,i]}(\tau)   \right] d\tau .
\end{split}
\end{equation}
We analyze the case presented in Prop.~\ref{prop:stealthRep}, 
\textcolor{ajg}{where the} attack is stealthy in the absence of the watermark.  
\textcolor{ajg}{It} is possible to formulate the following detectability condition:
\begin{proposition}\label{prop:detect}
	If, for some $t = T_d\geq T_a$, the \textcolor{ajg}{inequality}
	\begin{equation}\label{eq:detect}
	\left|S_j\delta_{[j,i]}(t) - \int_{T_a}^t e^{F_j(t-\tau)}\left[{\widehat{K}_j}\delta_{[j,i]}(\tau)\right]{d\tau}\right| > 2\bar{r}_{[j,i]}(t)
	\end{equation}
	\textcolor{ajg}{holds,} then detection of the replay attack is guaranteed. \textcolor{tp}{$\hfill \square$}
\end{proposition}
\begin{pf}
\textcolor{new}{	To guarantee detection, it is necessary for \eqref{eq:atkDetTst} to be violated, i.e. $|r_{[j,i]}(t)|>\bar{r}_{[j,i]}(t)$. Applying the triangle inequality to the explicit solution of $r_{[j,i]}(t) = {\epsilon}_{[j,i]}^a(t) + \rho_{[j]}(t-nT)+{\delta}_{[j,i]}(t)$, and ${\epsilon}_{[j,i]}^a(t)$ defined as in \eqref{eq:errRepAtk}, we have:
	\begin{equation}\label{eq:residualexplicit}
	\begin{split}
	|r&_{[j,i]}(t)| = \\
	= &{\left|{\epsilon}_{[j,i]}^a(t)+\rho_{[j]}(t-T)+\delta_{[j,i]}(t)\right|} + \\
	&{+ \left|e^{F_j(t-T_a)}\left(e_{[j,i]}^a(T_a) + H_j \rho_{[j]}(T_a-T)\right)+\right.}\\
	&{\left.- H_j\left(\rho_{[j]}(t-T)+\delta_{[j,i]}(t)\right) + \int_{T_a}^t e^{F_j(t-\tau)} \right.}\\
	&{\left.\left[S_j{w}_{[j]}(\tau-T) + (S_jB_jK_j-\widehat{K}_j)\rho_{[j]}(\tau-T) +\right.\right.}\\
	&{\left.\left.  - \widehat{K}_j {\delta}_{[j,i]}(\tau)   \right] d\tau + \rho_{[j]}(t-T) + \delta_{[j,i]}(t)\right|}\\
	\geq &
	\left|S_j\delta_{[j,i]}(t) - \int_{T_a}^t e^{F_j(t-\tau)}\left[{\widehat{K}_j}\delta_{[j,i]}(\tau)\right]{d\tau}\right| + \\
	& - \left| e^{F_j(t-T_a)}\left(e_{[j,i]}^a(T_a)+H_j\rho_{[j]}(T_a-T) \right) + \right.\\
	& + {S}_j\rho_{[j]}(t-T) + \int_{T_a}^t e^{F_j(t-\tau)} \left[ S_jw_{[j]}(\tau-T) + \right.\\
	&\left.\left. + \left(S_jB_jK_j-\widehat{K}_j\right)\rho_{[j]}(\tau-T)\right]d\tau\right|\\
	\geq & \left|S_j\delta_{[j,i]}(t) - \int_{T_a}^t e^{F_j(t-\tau)}\left[{\widehat{K}_j}\delta_{[j,i]}(\tau)\right]{d\tau}\right| - \bar{r}_{[j,i]}(t) {.}
	\end{split}
	\end{equation}
The last inequality derives from the bound on the disturbances, as well as from the hypothesis of Prop.~\ref{prop:stealthRep}, as can be seen in \eqref{eq:estErrBoundTa}. 
{From \eqref{eq:residualexplicit} and \eqref{eq:atkDetTst} we obtain}
	\begin{equation}
	\begin{split}
	|r&_{[j,i]}(t)| \geq \\
	\geq & \left|S_j\delta_{[j,i]}(t) - \int_{T_a}^t e^{F_j(t-\tau)}\left[{\widehat{K}_j}\delta_{[j,i]}(\tau)\right]{d\tau}\right| - \bar{r}_{[j,i]}(t) \\
	> &\quad \bar{r}_{[j,i]}(t),
	\end{split}
	\end{equation}
{which is the attack detection condition}. $\hfill\blacksquare$}
\end{pf}

\subsection{Watermark design}

\textcolor{ajg}{\textcolor{tp}{Now, we design a watermark signal}
$\Delta_{[j,i]}(t)$, \textcolor{tp}{that allows the detection of a replay attack through \eqref{eq:detect}, with the additional aim of}
obstructing the attacker from identifying the watermark from analysis of the communicated signal \eqref{eq:yCommWtmk}. \textcolor{tp}{In qualitative terms, we} note that these two objectives conflict with each other, as the first would benefit from large amplitude changes over time, which may, however, aid the identification of the watermark. Furthermore, we notice that the watermark enters condition in \eqref{eq:detect} through $\delta_{[j,i]}(t)$, rather than $\Delta_{[j,i]}(t)$ itself, \textcolor{tp}{thus} making it challenging to design a watermark satisfying \eqref{eq:detect} independently of $T_a$ and $T$, which are unknown to all but the attacker.}
For
\textcolor{new}{the purpose of}
this work, we introduce the following assumption, to simplify watermark design.
\begin{assumption}\label{ass:boundT}
	The attack period is upper bounded by some known quantity, i.e. $T\leq \bar{T}$. \textcolor{tp}{$\hfill \triangle$}
\end{assumption}

To justify that this assumption is mild, \textcolor{new}{we first note that if the attacker records data while the DGUs are in steady state,}
the attack to be stealthy to the \textcolor{ajg}{considered} monitoring scheme 
following Prop.~\ref{prop:stealthRep}. We also stress that we are considering a network of DGUs, whose steady state is determined by \textcolor{new}{the level of }
the load currents in all subsystems. It is
\textcolor{gft}{usually} possible to define an upper bound $\bar{T}$ as the maximum period between load changes within the DCmG, which can be evaluated empirically, or estimated \textit{a priori}.

Using this assumption, we then propose a sawtooth signal of period $2\bar{T}$ as the watermark:
\begin{equation} \label{eq:Delta}
\Delta_{[j,i]}(t) = c_{[j,i]}(t-2\nu\bar{T}) ,
\end{equation}
where $\nu = \left\lfloor {t}/{2\bar{T}}\right\rfloor$, and $c_{[j,i]}$ is the slope. The period $2\bar{T}$ is selected to avoid having $\delta_{[j,i]}(t)=0$ for any ${T \in (0,\bar{T}]}$. Indeed, with this 
\textcolor{new}{choice of the}
watermark signal, $\delta_{[j,i]}(t)$ will be a square wave of period $2\bar{T}$.




\textcolor{ajg}{In the following we show through simulation how it is possible to tune the proposed watermark such that it is difficult to identify by the attacker, while nonetheless enabling detection of a replay attack which is stealthy for the monitoring scheme proposed in \citep{gallo2018distributed}.}

\begin{figure}
	\centering
	\includegraphics[width=\linewidth] {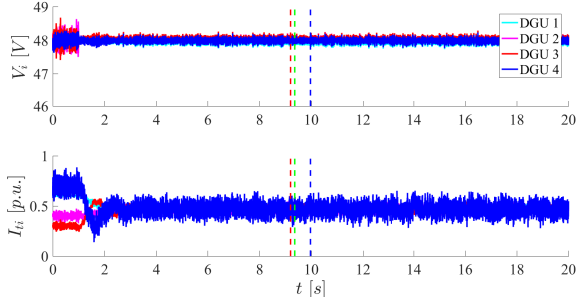}
	\caption{Voltage and current states from DGUs 1-4. \textcolor{ajg}{Vertical lines show initial time of attack on DGU~$4$ (red), and detection of replay attack in communication with DGU~$2$~and~$3$ (in blue and green, respectively).}
	}
	\label{fig:states_w_wm}
\end{figure}

\section{Simulation Results}\label{ch:Sim}

The proposed detection scheme with watermark is \textcolor{ajg}{validated through} 
simulations in MATLAB. 
A DC microgrid consisting of 4 DGUs, \textcolor{ajg}{interconnected} 
as in Fig~\ref{fig:mG_Struct}, is considered, where the parameters and matrices for DGUs and UIOs are taken as in \citep{gallo2018distributed}. Process and measurement noises are 
drawn from uncorrelated uniform distributions satisfying Assumption \ref{ass:Bounds}, where 
${\bar{w}_{[i]} = \left[
	0.1,
	0.1,
	0.1
	\right]^\top}$ and
${\bar{\rho}_{[i]} = \left[
	0.01,
	0.01,
	0.01
	\right]^\top}.$
\textcolor{gft}{These noises induce variations in the electrical signals comparable to those seen in real microgrids, due to the converter operations and measurement noises.}

Before time $t=1s$, all the DGUs are disconnected and 
\textcolor{gft}{do not communicate}. At time $t=1s$, neighboring DGUs connect to each other to create the microgrid topology given in Fig~\ref{fig:mG_Struct}. The current loads at the PCCs of DGUs 1, 2, and 3 are considered to periodically change from $6A$ to $6.2A$, $4A$ to $4.25A$, and $3A$ to $3.15A$, respectively, to show the dynamic nature of loads in a microgrid. The attacker chooses $T_a = 9.2s$ to start replaying the data it recorded starting at $t = 7.4s$, i.e. choosing attack period $T = \textcolor{ajg}{\bar{T} =} 1.8s$\textcolor{ajg}{, so that the data is recorded in steady state.}

Each DGU adds a watermark signal as in \eqref{eq:Delta} to its measurements communicated to its neighbors. \textcolor{ajg}{Added watermarks are the same for each communication link,}
i.e., $c_{[i,j]}=c_{[i,k]} \enskip \forall i \in \mathcal{N}, \forall j,k \in \mathcal{N}_i$, 
\textcolor{ajg}{to prevent attacker from identifying the watermark from the differences between outgoing communication from the same DGU.}
The slopes $c_{[j,i]}$ of the watermarks take constant values from $10^{-3.2}$ to $10^{-3.5}$ for each DGU.

\begin{figure}
	\centering
	\includegraphics[width=\linewidth]{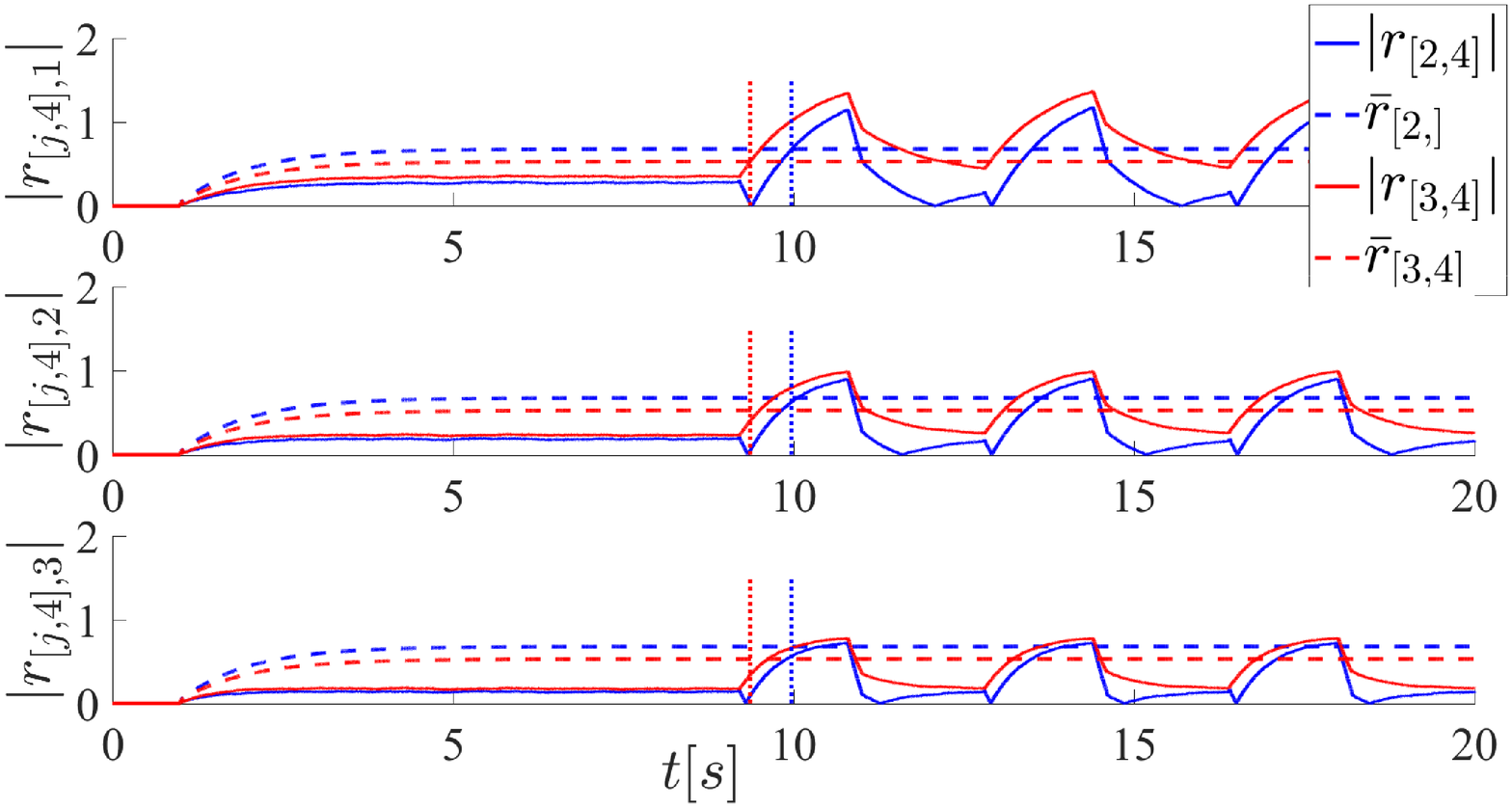}
	\caption{\textcolor{ajg}{Comparison of residuals and thresholds, in solid and dotted lines, respectively, of monitors in DGU~$4$ with watermarking. Data regarding estimation of state of DGU~$2$ is presented in blue, while data referring to DGU~$3$ is in red. Detection occurs as one component of the residual exceeds its threshold (vertical lines).}
	}
	\label{fig:res_w_wm}
\end{figure}

\textcolor{ajg}{Voltages and currents of each DGU of the network are shown in Fig.~\ref{fig:states_w_wm}, whereas Fig.~\ref{fig:res_w_wm} shows the comparison of the residuals of DGU~$4$'s estimates of its neighbors, compared to their respective thresholds. It can be seen that detection occurs for both communicated measurements, although the effect of the replay attack on the DCmG is not noticeable from Fig.~\ref{fig:states_w_wm}.} The relatively late detection based on the residual of the UIO estimating the states of DGU 3 is due to the smaller watermark slope $c_{[3,i]}$.
\textcolor{ajg}{We also show, Fig.~\ref{fig:res_wo_wm}, that if the watermark were not present, the detection of the attack would not occur.}

\begin{figure}
	\centering
	\includegraphics[width=\linewidth]{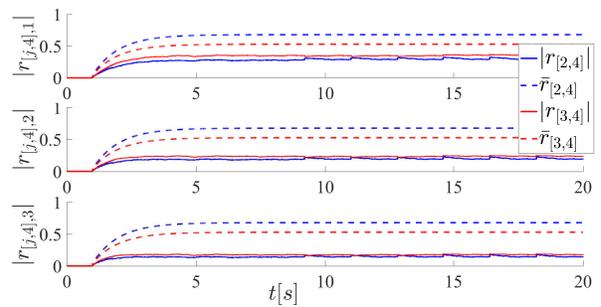}
	\caption{\textcolor{ajg}{Comparison of residuals and thresholds of estimators in DGU~$4$ without watermark (colors used as in Fig.~$3$). As can be seen, the attack is not detected. }
	}
	\label{fig:res_wo_wm}
\end{figure}

\textcolor{ajg}{\textit{Enabling Detection:\hspace{.1cm}} We now provide some insight regarding the design of the watermark. Specifically, we note that in \eqref{eq:errRepAtk}, while $w_{[j]}$ and $\rho_{[j]}$ are multiplied by $S_j$ and $(S_jK_jB_j-\widehat{K}_j)$, respectively, $\delta_{[j,i]}(t)$ is scaled by $\widehat{K}_j$. This is a design parameter of the UIO which, as seen in \citep{gallo2018distributed}, can be made to have a larger absolute value than the other scaling matrices. Therefore, even with $c_{[j,i]}$ small, $\Delta_{[j,i]}(t)$ will have a large impact on $r_{[j,i]}$ when the microgrid is subject to attack.}


\textit{Identifiability of the watermark:\hspace{.1cm}} We now focus on the identifiability of the watermark by the attacker from the communicated data. The methods which have been considered possible to be used by the attacker are: by inspection, \textcolor{new}{via distribution and statistics,} and through frequency spectrum analysis \textcolor{new}{of communicated data}. In terms of identifiability by inspection, \textcolor{mst}{we note that $y^c_{[j,i]} = x_{[j]} + \rho_{[j]} + \Delta_{[j,i]}$, and that the maximum value of a watermark signal over the whole microgrid, $2\bar{T}10^{-3.2}$, is less than a quarter of the bound on the measurement noise of the corresponding DGU,} and hence the discontinuity occurring every period will be masked by noise. 

\textcolor{new}{Next, we show the distribution of the first component of the communicated and actual outputs of DGU~1 in Fig.~\ref{fig:histogram}. It is observed from this figure that \textcolor{ajg}{the addition of the watermark does not cause any significant changes to the distribution of the communicated measurements, compared to the measurement output themselves.}
Furthermore, the addition of watermark on the output signals changed the mean of the measurements of all DGUs by at most $0.45\%$.
Moreover, differences in the variances of these signals were below $3\%$ of the original value for all DGUs. Therefore, the watermark signal is difficult to identify from the distribution and the statistics of the communicated outputs. 
}

\begin{figure}
	\centering
	\includegraphics[width=\linewidth]{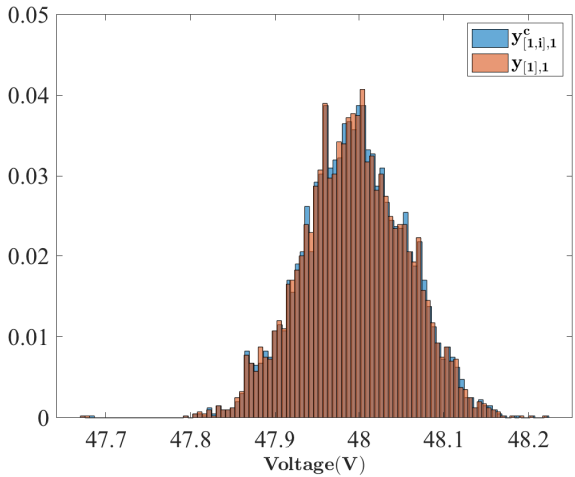}
	\caption{ Comparison of the distributions of the communicated output (in blue), actual output (in red), of the voltage state of DGU 1 from $t=5s$ to $t=9s$. 
	}
	\label{fig:histogram}
\end{figure}

\textcolor{new}{Finally}, if \textcolor{mst}{the watermark causes significant changes in the frequency domain characteristics of the communicated outputs, it may be possible for the attacker to identify it.} Hence, in Fig.~\ref{fig:ffts_w_wm} we compare the frequency domain of the measurements of $v_{[1]}$ with and without the watermark, by analyzing their fast Fourier transforms (FFTs) for $t \in [5s,9s]$ 
{(i.e.~when the system is almost in steady state, which is favorable for watermark detection)}, as well as comparing them to that of the watermark itself. \textcolor{mst}{One can see from this figure that the FFTs of the communicated and actual outputs are very similar and that of the watermark signal is incomparably small. \textcolor{new}{As the watermark is periodical, one would expect there to be a peak in the spectrum at its frequency. In fact, $f_{\Delta} = \frac{1}{2\bar{T}} = 0.2778 Hz$, and at this frequency there is a small peak. We note, however, that, as the main frequency spectrum of the measured and communicated outputs for the DGUs lies at low frequencies, the peak caused by the watermark is masked, as the spectral density of the communicated signals are larger.}
Hence we stipulate that it is difficult for the attacker to identify the watermark signal from \textcolor{gft}{a spectral analysis of} the communicated outputs.}

\section{Conclusions}\label{ch:Concl}
\textcolor{ajg}{In this work we have presented a distributed watermarking strategy to support a monitoring scheme used to validate information transmitted between DGUs of a DCmG. Analysis of the monitoring scheme under replay attack, without the additive watermark, shows that as long as data is recorded in steady state, the attack is undetectable. We then introduce the watermark, and we derive a condition on the watermark to guarantee detection. Finally, we propose a preliminary watermark signal design, showing its effectiveness through simulation.}
As future work, we \textcolor{fb}{plan to consider non-ideal communication networks,}
and to propose a more refined watermarking signal.

\begin{figure}
	\centering
	\includegraphics[width=\linewidth]{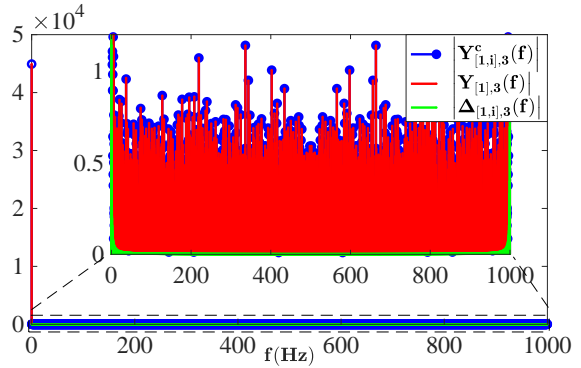}
	\caption{ Comparison of FFTs of the communicated output (in blue), actual output (in red), and the added watermark (in green) of the integrator state of DGU 1 from $t=5s$ to $t=9s$. 
	}
	\label{fig:ffts_w_wm}
\end{figure}

\textcolor{new}{\appendix
\section{DGU Dynamics}\label{app:DGU}
The voltage and current dynamics for DGU~$i$ are characterised as follows:
\begin{equation}
\begin{split}
&\frac{dV_i}{dt} = \frac{1}{C_{ti}}I_{ti} + \sum_{j \in \NN_i}\frac{1}{C_{ti}R_{ij}}\left(V_j-V_i\right) - \frac{1}{C_{ti}}I_{Li} \\
&\frac{dI_{ti}}{dt} = \frac{1}{L_{ti}}V_{ti}  - \frac{R_{ti}}{L_{ti}}I_{ti} - \frac{1}{L_{ti}}V_{i}
\end{split},
\end{equation}
where $(V_{ti},I_{Li})$ are inputs to the DGU, $(V_i,I_{ti})$ are the states, $V_j\in\NN_i$ are the interconnection terms between the DGUs. $R_{ti}$, $C_{ti}$, $L_{ti}$ are electrical parameters of the RLC filter of DGU~$i$. $R_{ij}$ is the resistance of the power line connecting DGUs $i$ and $j$.
Matrices $A_{ii}$, $B_i$, $M_i$, $A_{ij}$, $K_i$, and $C_i$ in \eqref{eq:DGUi} are defined as \citep{tucci2016decentralized}:
\small
\[
A_{ii} = \left[
\begin{array}{ccc}
-\sum_{j \in \NN_i} \frac{1}{R_{ij}C_{ti}} &\frac{1}{C_{ti}}       &0 \\
-\frac{1}{L_{ti}}                          &-\frac{R_{ti}}{L_{ti}} &0 \\
-1                                         &0                      &0 
\end{array}
\right],\]
\[
B_i = \left[
\begin{array}{c}
0\\
\frac{1}{L_{ti}}\\
0
\end{array}
\right],
K_i = \left[
\begin{array}{ccc}
k_{i,1} &k_{i,2} &k_{i,3}
\end{array}
\right],
\]
\[
M_i = \left[
\begin{array}{cc}
-\frac{1}{C_{ti}} &0 \\
0                 &0 \\
0                 &1
\end{array}
\right],
A_{ij} = \left[
\begin{array}{ccc}
\frac{1}{R_{ij}C_{ti}}  &0 &0 \\
0  &0  &0 \\
0  &0  &0 
\end{array}
\right],
C_i = I,
\]
\normalsize 
where $I$ is the identity, and $G_i$ is the second column of $M_i$, i.e. $G_i = [0,0,1]^\top$.}

\bibliography{NecSys18_RepAtk,biblio_22_1_2017}

\end{document}